\documentstyle[prl,aps,twocolumn]{revtex}
\begin{document}
\twocolumn[\hsize\textwidth\columnwidth\hsize\csname
@twocolumnfalse\endcsname
\draft
\title{Concentrating Entanglement by Local Actions---Beyond  Mean Values}
\author{Hoi-Kwong Lo}
\address{Hewlett-Packard Labs,
Filton Road, Stoke Gifford, Bristol, UK, BS34 8QZ}
\author{and}
\author{Sandu Popescu}
\address{Isaac Newton Institute, 20 Clarkson Road,
Cambridge, UK,  CB3 0EH}
\address{and}
\address{BRIMS, Hewlett-Packard Labs,
Filton Road, Stoke Gifford, Bristol, UK, BS34 8QZ}

\date{\today}
\preprint{quant-ph:}
\maketitle
\begin{abstract}
Suppose two distant observers Alice and Bob share
a pure bipartite
quantum state. By applying local operations and communicating with each
other using a classical channel, Alice and Bob can
manipulate it into some other states.
Previous investigations of entanglement manipulations
have been largely
limited to a small number of strategies and their 
average outcomes. Here we consider a general
entanglement manipulation strategy and go beyond the average property.
For a {\it pure} entangled state shared between
two separated persons Alice and Bob, we show that
the {\it mathematical} interchange symmetry of the Schmidt
decomposition can be promoted into a {\it physical} symmetry
between the actions of Alice and Bob. Consequently, the most general
(multi-step two-way-communications) strategy of entanglement
manipulation of a pure state is, in fact, equivalent to a strategy
involving only a single (generalized) measurement by Alice
followed by one-way communications of its result
to Bob. We also prove that strategies with one-way communications
are generally more powerful than those without communications.
In summary, one-way communications is necessary and sufficient
for the entanglement manipulations of a {\it pure} bipartite state.
The supremum
probability of obtaining a maximally entangled state (of any
dimension) from an arbitrary state is determined and a strategy
for achieving this probability is constructed explicitly.
One important question is whether collective
manipulations in quantum mechanics
can greatly enhance the probability of large deviations
from the average behavior. We answer this question
in the negative by showing that, given $n$ pairs of identical
partly entangled {\it pure} states $| \Psi \rangle) $ with entropy of
entanglement $E( | \Psi \rangle) $,
the probability of getting $n K $ ($K >  E( | \Psi \rangle) $)
singlets out of entanglement concentration tends to
zero as $n$ tends to infinity.

\end{abstract}
]
\narrowtext

\section{Introduction And Summary of Key Results}
\label{Intro}

Entanglement---the non-locality of entangled state---as in the
Einstein-Podolsky-Rosen paradox\cite{EPR},
discovered by J. Bell\cite{Bell} in 1964, has
long been regarded a hallmark of quantum
mechanics. In the past, entanglement was often regarded as a
qualitative property of a state. The last few years,
however, witnessed a dramatic change in the
approach to entanglement. Entanglement is now regarded
as an important quantitative and useful resource in achieving tasks
of quantum information processing such as dense coding\cite{dense},
teleportation\cite{tele} and reduction of communication
complexity\cite{complex}.

The study of quantum information processing
has been complicated by the fact that entanglement can appear in
many non-standard forms.
Fortunately, it is known that
two distant parties sharing a bipartite pure
state can apply local operations and
classical communication to ``manipulate'' entanglement,
thus converting one form to another\cite{BBPS}.
To better understand quantum information processing,
it is important to discover the
fundmental laws of a general
entanglement manipulation. Most of the previous investigations
have focused on some particular
types of entanglement manipulations, namely, entanglement
concentration and dilution of an ensemble of identical states,
say $\Omega = ( a | 11 \rangle + b | 22 \rangle)^N$, which
are collectively processed.
Moreover, the main interest was in the average properties and
little is known about the actual probability
distribution of the outcomes of those manipulations.

This paper concerns mainly the fundamental laws of
entanglement manipulations of {\it pure} bipartite states.
Unlike previous investigations,
here we allow the initial state to be a single copy of a
general state $\Psi$.
It is useful to note, however, that in fact all entanglement manipulation 
methods, both ``single-pair'' and 
``collective'' ones can be reformulated as
``single-pair'' methods,
by redefining 
the ``particles''. Indeed, suppose Alice and Bob share $n$ pairs of particles, 
and intend to process them by some collective method. We can now regard 
all $n$ 
particles in each side as a single ``particle'', living in a higher 
dimensional Hilbert space (equal to the product of the Hilbert spaces of the 
original $n$ particles). The $n$ original pairs can thus be regarded a single 
pair of two (more complex) quantum particles, and the original ``collective'' 
manipulation can be regarded as a ``single-pair'' type manipulation
of this new 
pair. Consequently, all questions concerning collective manipulations
can be
answered by studying ``single-pair'' manipulations of a generic state of  
two arbitrary particles. This is the path that we will follow in the present
paper.

Our results are the following.

\begin{itemize}

\item
First of
all, we show that, rather surprisingly,
general entanglement manipulations of
a pure bipartite state with only one way communication
are equally powerful as those with two way communication,
but are more powerful than those with no communication.

\item
Then, we specialize in a class of entanglement
manipulations, namely entanglement concentration, from
a general pure bipartite initial state $\Psi$ to a $m$-dimensionally
maximally entangled state (which we shall call $m$-$ME$-state),
$\Phi_m = { 1 \over \sqrt{m} } \sum_{i =1}^m
| i \rangle_A | i \rangle_B $ where $| i \rangle_A$ and
$| i \rangle_B $ are orthonormal vectors in Alice's and Bob's
Hilbert
spaces $H_A$ and $H_B$ respectively and $m$ is fixed but
arbitrary. (Notice that $\Phi_1$ is a direct product, $\Phi_2$ is
a singlet and $\Phi_{2^q}$ is equivalent to $q$ singlets.
Therefore, a $m$-$ME$-state can be regarded as a generalization
of singlets.) The main questions that we ask are:
What is the optimal probability of
getting a $\Phi_m$ from $\Psi$? and also:
What is the optimal strategy that will achieve
such a probability? In this paper, we give complete
answers to those questions.

\item
After that, we specialize to
the entanglement manipulation of
a large number of identical pairs of $\Phi$ and
derive a bound to the probability of having
a large deviation from the average property.
More specifically, suppose
that two remote observers, Alice and Bob, share n pairs of
spin 1/2 particles, each pair in a non-maximally entangled pure state
$|\Psi\rangle=\alpha |1\rangle |1\rangle+\beta |2\rangle|2\rangle$. 
Then, by 
local actions (which may include local unitary transformations,
measurements and attachment of ancillary quantum systems)
and classical communications
Alice and Bob can convert these pairs into a (smaller) number $m$ of perfect
singlets.  It has been shown\cite{BBPS}
that in the limit of large $n$, Alice and Bob
can perform a {\it reversible } conversion of the $n$ pairs
$\Psi$ into singlets, obtaining, {\it on average} a number
$\bar m=nE(\Psi)$ of
singlets. (Here, $E(\Psi)$ is the ``entropy of
entanglement'' of a pure bipartite state consisting of
subsystems $A$ and $B$
and is defined to be the von Neumman entropy of subsystem A
(or B)~\cite{BBPS}.)
Furthermore, as a
consequence of this reversibility property, together with the fact that on
average entanglement cannot increase via local actions and classical
communications\cite{BBPS,BDSW}, this particular entanglement
manipulation method yields the maximal possible average
number of singlets. It has been shown \cite{Note0} that 
$E(\Psi)$, the maximal average number of singlets
which can be extracted 
per
original pair $\Psi$, is up to a constant multiplicative
factor, the {\it unique} measure of entanglement for
$\Psi$ that is additive and non-increasing under local
operations and classical communication.

However, in all previous investigations of entanglement manipulations,
the focus was on the
{\it average values}, such as on the question
``What is the average number of
singlets which can be extracted from n pairs $\Psi$?"
Indeed, the whole idea of reversibility refers only to
the average properties. The point is that, as in all asymptotic results,
there is always a very small but non-zero probability for
an entanglement concentration procedure to give a
substantially smaller amount of singlets than the expected number.
Here we want to go beyond
average values and ask about the actual distributions.
For example, the same
average number of singlets, $\bar m=nE(\Psi)$ might,
in principle, be obtained
from
very different distributions:  In the reversible procedure described in
Ref.\ \cite{BBPS},
out
of n pairs $\Psi$ a number m of singlets is obtained with some
probability $P_m$, and the distribution is essentially
Gaussian, peaked around $\bar m=nE(\Psi)$.  In particular,
via this procedure the probability to obtain a large number of singlets,
$m\approx n$ is exponential small. However,
one could envisage a distribution which
yields the same average $\bar m=nE(\Psi)$ while having a
non-negligible probability for
obtaining a large number of singlets---for example,
a distribution in which the
probability of obtaining $m=n$ singlets is $E(\Psi)$
while in all other cases
zero singlets are obtained.  The question is ``Does there
exist any entanglement
manipulation procedure which realizes the later distribution?"

A main point of our investigation is to gain a better understanding of the
{\it collective properties} involved in entanglement manipulation.  Indeed, if
Alice and Bob would extract singlets by processing each of the $n$ pairs $\Psi$
separately, the law of large numbers tells that the probability distribution of
the number of singlets will (asymptotically) be Gaussian. Deviations from this 
distribution can be obtained (if at all) only if Alice and Bob process all the 
$n$ pairs together. But are such deviations possible? And if so, how big 
can they be?

[To put things in the right perspective, we would like to mention that the  
reversible procedure\cite{BBPS}
discussed above, is {\it not} a procedure
in which each 
pair is processed separately but a collective one---yet, the distribution it 
yields is essentially Gaussian.]

In this paper, we show that, in the context of entanglement
manipulations of a
large ensemble of identical pure bipartite states,
collective
manipulations cannot
substantially enhance the probability of large deviations
from the average properties.

\item
Afterwards, we consider a subclass of entanglement
manipulation strategies in which the final state is always
one of the possible $\Phi_m$'s. (Say, the state
$\Phi_1$ appears with
probability $p_1$, $\Phi_2$ with probability
$p_2$ and, ...., $\Phi_n$ with probability $p_n$, etc)
We define a natural notion of a universal strategy for
entanglement manipulations for all values of $m$ and
show that such a universal strategy cannot possibly
exist.

\item
Finally, we present open questions in the entanglement
manipulations of pure states and
discuss briefly problems in generalizing our results to
mixed states.

\end{itemize}

In summary, in a conceptual level, the novelties of
our investigation are the following.

1) Some of our results---particularly the
statement that manipulations with
one way communication are equally powerful as those with
two way communication---apply to a general
entanglement manipulation strategy. In contrast,
all previous investigations have focused on either entanglement
concentration or dilution.

Incidentally, we clearly
demonstrate the important role of symmetry in
entanglement manipulations. Indeed,
in our proof that manipulation strategies of a pure bipartite state
with only
one way communication are equally powerful as those with
two way communication, we are essentially promoting the
mathematical interchange symmetry of the
Schmidt decomposition into a physical symmetry between
the local actions of Alice and Bob. (See section~\ref{reduction}.)

2) We allow the initial state to be a general $\Psi$.
In contrast, almost all previous investigations have restricted
$\Phi$ to be $N$ identical copies of some state, say
$u= a | 11 \rangle + b | 22 \rangle $,
and have been mainly concerned with properties
in the
large $N$ limit.

3) We are concerned with the role of probability rather than
the average property in entanglement manipulations.

\section{One-way communications is necessary and sufficient for
entanglement manipulations of bipartite pure states}
\label{reduction}
\subsection{Reduction from two-way to one-way communications}

The most general scheme
of entanglement manipulations of a bipartite
pure entangled state involves two-way communications
between Alice and Bob. It goes as follows: Alice
performs a measurement and tells Bob the outcome.
Bob then performs a measurement (the type of measurement that
Bob performs can depend on Alice's measurement outcome) and
tells Alice the outcome, etc, etc.
The goal of this subsection is to prove
that any strategy of entanglement manipulation of
a pure bipartite state is equivalent to a strategy involving only
a {\it single} (generalized) measurement by Alice
followed by the {\it one-way} communications of
the result from Alice to Bob (and finally local unitary transformations
by Alice and Bob).

Here we introduce some definition.

{\bf Definition~1}: (Ordered Schmidt coefficients)
An arbitrary pure state $\Psi$ can be written in Schmidt
decomposition\cite{Appen}
\begin{equation}
\Psi=\sum_i^N \sqrt {\lambda_i}|a_i\rangle |b_i\rangle,
\label{eq1}
\end{equation}
where $ \langle a_i | a_j \rangle = 
\langle b_i | b_j \rangle = \delta_{ij}$.
We call $ \sqrt {\lambda_i}$'s the {\it ordered Schmidt coefficients}
if the $\lambda_i$'s  are ordered decreasingly, i.e.,
$\lambda_1 \geq \lambda_2 \geq \cdots \geq \lambda_N$.
Note that all phases have been absorbed in the
definition of the
states $|a_i\rangle$'s
so that the $\lambda_i$'s are positive real numbers.

First of all, since it is more convenient to deal with
projection operators than positive operator valued measures (POVMs),
we include any ancilla (measuring apparatus)
in Alice and Bob's quantum systems. Therefore,
without loss of generality,
we regard Alice and Bob as
sharing a pair of particles with an infinite (or an arbitrarily large)
dimensional Hilbert space but initially
only $N$ of the coefficients of the Schmidt
decomposition\cite{Appen} are non-zero,
i.e., $| \Psi \rangle = \sum_{i=1}^N \sqrt{ \lambda_i}
| a_i \rangle |b_i \rangle$
where $  \langle a_i | a_j \rangle = \delta_{ij}$
and $\langle b_i | b_j \rangle = \delta_{ij}$.
We further assume that the above form of the Schmidt decomposition of
$| \Psi \rangle$ is known
to Alice and Bob.

Second, we consider only the
most advantageous entanglement manipulation scheme in each step of which
Alice keeps track of the results of all her measurements
and tells Bob about them and vice versa.
Alice and Bob then update their information on
the state they share in each step.
Since it is a pure state $| \Psi \rangle$ that Alice
and Bob start with,
they always deal with a {\it pure state} in {\it each step}. 
Any scheme
in which Alice and Bob choose to be sloppy or ignorant can be
re-casted as a situation in which they fail to
make full use of their information. Therefore,
there is no loss in generality in our consideration.\cite{addremark1}

We now argue
that any two-way entanglement manipulation strategy
for the state $| \Psi \rangle$ can be
re-casted into an equivalent strategy which
involves only one-way communications from Alice to Bob---that
is to say a strategy
in which Alice performs all the measurements and informs Bob of
the outcomes afterwards.
This is so because (i) in entanglement manipulations
we are mainly concerned with the coefficients of the
Schmidt decomposition and (ii) in {\it each} step of entanglement manipulation,
the Schmidt decomposition of the {\it pure} state involved is
always {\it symmetric} under the interchange of Alice and Bob.
With such symmetry, there is no advantage in having
Bob perform the measurement instead of Alice\cite{extension,sym}.

More concretely, consider a round of
communications
in a two-way scheme of
entanglement manipulation. Suppose Alice has performed a measurement on
$| \Psi'  \rangle = \sum_k \sqrt{ \lambda'_k }| a'_k \rangle | b'_k \rangle $
and
obtained an outcome $o_1$.
She can work out the Schmidt decomposition $ P_{o_1}| \Psi' \rangle =
\sum_k \sqrt{\lambda''_k} | a''_k \rangle | b''_k \rangle$ of
the state that she now shares with Bob.
Now Alice is supposed to tell Bob the outcome
$o_1$ of her measurement and Bob then will perform a
measurement with a set of local
projection operators say $\{ P^{Bob}_l\}$.
However, it turns out that there exists a set of
local projection operators $\{ P^{Alice}_l\}$ by Alice which will
do essentially the same trick, as far as entanglement
manipulation is concerned.
Mathematically, we claim the following Proposition.

{\bf Proposition~1}: 
Given any pure bipartite
state $| \Psi \rangle_{AB}$ shared by Alice and Bob
and any complete
set of projection operators $\{ P^{Bob}_l\}$'s by Bob,
there exists a complete set of projection operators
$\{ P^{Alice}_l\}$'s by Alice and,
for each outcome $l$, a direct product of
local unitary transformations $U^A_l \otimes U^B_l$ such that,
for each $l$,
\begin{equation}
\left( I \otimes P^{\rm Bob}_l \right)| \Psi \rangle
= \left( U^A_l \otimes U^B_l \right) \left( P^{\rm Alice }_l \otimes I 
\right) | \Psi \rangle .
\label{symmetry}
\end{equation}

{\it Idea of the Proof}:
Consider a pure bipartite state in its Schmidt decomposition
\begin{equation}
| \Psi \rangle  = \sum_ i \sqrt{ \rho_i} | e_i \rangle_A | e_i \rangle_B
\end{equation}
shared between Alice and Bob.
Notice that, by the definition of the Schmidt decomposition,
it is symmetric
under the interchange of $| e_i \rangle_A $ and $ | e_i \rangle_B$.
This is a mathematical symmetry. Now, in Proposition~1, we
promote this mathematical symmetry into a physical symmetry
between the actions of Alice and Bob in the context of
entanglement manipulations.
More precisely, if Bob applies a set of projection operators
$\{ P^{Bob}_l\}$ on his system and obtains an outcome $l$,
the state $| \Psi \rangle$
will be transformed into some state, say,
\begin{equation}
| \Psi^B \rangle  
         = \sum_ i \sqrt{ \mu_i} | a'_i \rangle_A
| b'_i \rangle_B . 
\end{equation}
On the other hand, if Alice applies a corresponding
set of projector operators $\{ P^{Alice}_l\}$ on her system
(instead of Bob), then we show that the corresponding outcome $l$
will give her a state
\begin{equation}
|  \Psi^A \rangle  
         = \sum_ i \sqrt{ \mu_i} | a''_i \rangle_A
| b''_i \rangle_B  
\end{equation}
with exactly the same Schmidt coefficients as $| \Psi^B  \rangle$.
Consequently, there exists a bi-local unitary transformation
that will rotate the state $| \Psi^A \rangle  $ to
$| \Psi^B  \rangle $. In this sense, the states
$| \Psi^A \rangle  $ and
$| \Psi^B  \rangle $
are equivalent. The upshot is that there is no advantage for
Bob to perform a measurement, in the place of Alice.
In summary, as far as entanglement manipulation of a pure bipartite
state is
concerned, there is a total symmetry
between the actions of Alice and Bob\cite{invariant}.

{\bf Proof}: See Appendix~\ref{a:proofprop1}.

One can repeat the above argument and prove that all the rounds
of measurements
can be performed by Alice alone and Alice only needs to tell Bob
her outcomes after the completion of all her measurements.

What it means is that, for Alice and Bob manipulating
a pure bipartite state,
one can, without loss of generality,
restrict oneself to schemes of 
entanglement manipulations using only one-way communications from Alice to
Bob.

Finally,  it is a well-known consequence of measurement
theory that the entire 
sequence of Alice's measurements can be described as a {\it single}
generalized  measurement.
[One may argue this well-known result as follows.
Every measurement consists of two steps---the interaction of
a measuring devise with a system, and the ``reading" of the
measuring device, i.e. a unitary transformation and a projection.
Now, any
arbitrary sequence of independent measurements can be replaced by an
equivalent single measurement, by simply letting all the interactions to
be performed first, and reading all the measuring devices simultaneously
at the end. In this case one can view all the independent measuring
devices as a (more complicated) {\it single} measuring device,
performing
a {\it single} interaction with the measured system
(the unitary transformation
describing this interaction being simply the product of the unitary
transformations describing the individual measuring devices)
and followed
by a {\it single} reading stage.
Furthermore, even if the measurements are not independent from each other,
i.e., some measurements depend on the results of previous
measurements, we
can still replace the sequence by a single measurement:
In this case too the human observer can postpone ``reading'' the
results obtained by the different measuring devices until
the end. Indeed, there is no need for the observer to read the results of
the measurements in order to tune the subsequent measurements accordingly.
The
entire process can be realized by the measuring devices interacting with
each other
as well as with the system under observation.
Then, once again, we have a single measuring device,
performing a single interaction, (only that the interactions between the
measuring device and the system contain also some internal interactions
between the different parts of the measuring device---corresponding to
one part reading the result of the other), and a single reading stage.]

In summary, the most general strategy of entanglement
manipulation of a pure bipartite
state is equivalent to a strategy involving
only a single (generalized) measurement performed by Alice
followed by the one-way communications of the result from Alice
to Bob
(and finally local unitary transformations by Alice and Bob).

\subsection{One-way communications are
provably better than no communications}
We have shown above that two-way communications are not
necessary for the entanglement manipulation of a pure
bipartite state---the most general
entanglement manipulation strategy can be realized with only
one-way communication. A natural question to ask is whether
communication is needed at all. We show that, indeed, communication is
necessary. That is to say that entanglement manipulation strategies
without communication cannot achieve all that could be achieved with
communication. The proof of relegated to Appendix~\ref{noway}.

In conclusion, one-way communications generally give
more powerful strategies than those without communications.
On the other hand, we proved in the above paragraphs
that one-way communications
is sufficient for any strategy. Combining these two results,
we conclude that one-way communications is necessary and
sufficient for implementing a general strategy of entanglement
manipulations of pure bipartite states.

\section{Obtaining a given maximally entangled state $\Phi_m$ from an arbitrary 
state $\Psi$}

{\bf Definition~2}: (m-ME-state: $\Phi_m$)
We shall denote by $\Phi_m$ a standard
{\it m-dimensional maximally entangled 
state} 
\begin{equation}
| \Phi_m \rangle = { 1 \over \sqrt{m} } \sum_{i=1}^m
| i \rangle_A | i \rangle_B, 
\label{eq2}
\end{equation}
where $| i \rangle_A$'s ($| i \rangle_B$'s respectively) 
form an orthonormal basis for a Hilbert space
$H_A$ ($H_B$ respectively).
In particular, $\Phi_1$ is a direct-product,
$\Phi_2$ is (equivalent to) a
singlet and $\Phi_{2^q}$ is equivalent to $q$ singlet pairs.
In what follows,
we shall call $\Phi_m$ an {\it m-ME-state}.

We now come to one of the main results of our paper.
We consider the following particular problem.
Suppose Alice and Bob share a pair of particles in 
some arbitrary pure state $\Psi$.  By different entanglement manipulations strategies 
we can transform $\Psi$ into a given m-dimensional maximally entangled state 
$\Phi_m$. In general such a process does not succeed with certainty but only 
with some probability $p_m$. Here we enquire what is the maximal probability 
with which such a transformation could occur.

Incidentally, one can even obtain a maximally entangled state whose degree of 
entanglement is {\it greater} than that of the initial state. Since the average 
degree of entanglement cannot increase, it is obvious that such a transformation 
always occurs with probability less than 1. To describe such a situation we 
sometime use the term ``gambling with entanglement" - indeed, Alice and Bob try 
to achieve a better than average outcome while taking the rise of loosing 
entanglement if the result turns out to be unfavorable.

{\bf Definition~3}: ($p_m^{MAX}$) For any positive integer $m$,
we define $p_m^{MAX}$\cite{max} to be the
supremum over all manipulation strategies of the 
probability $p_m$ of getting an m-state
$\Phi_m$ from a pair initially in the state $\Psi$.

We will prove the following theorem:
{\bf Main Theorem (Theorem~0)}: If we write the initial
state $\Psi $ in ordered Schmidt decomposition
as $\Psi= \sum_{i=1}^N \sqrt{\lambda_i}
| a_i \rangle | b_i \rangle$, 
the supremum probability $p_m^{MAX}$ of
obtaining $\Phi_m$ over all possible
entanglement manipulation strategies is
given by the following.

i) If $m>N$ (N being the number of terms in the Schmidt decomposition of
$\Psi$),
then  $p_m^{MAX}=0$.

ii) If $m\leq N$, then 

\begin{equation}
 p_m^{MAX}= min_{1 \leq r \leq m}  ~{ m \over r}
( \lambda_{m-r+1} + \lambda_{m-r+2} +
\cdots  + \lambda_N ).
\label{eq3}
\end{equation}

The proof of this main theorem is divided into two parts.
In the next section (section~\ref{s:bound}),
we will derive an upper bound on
$p^{MAX}_m$ (see Theorem~1). In section~\ref{s:optimals}, we
demonstrate an explicit strategy that
saturates the bound and is, thus, optimal. (See Theorem~2.)

\section{Upper Bound on $p^{MAX}_m$: Theorem 1}
\label{s:bound}
{\bf Theorem~1}:  If we write the initial
state $\Psi $ in ordered Schmidt decomposition
as $\Psi= \sum_{i=1}^N \sqrt{\lambda_i}
| a_i \rangle | b_i \rangle$, 
the supremum probability $p_m^{MAX}$ of
obtaining $\Phi_m$ over all possible
entanglement manipulation strategies
satisfies the following.

i) If $m>N$ (N being the number of terms in the Schmidt decomposition of
$\Psi$),
then  $p_m^{MAX}=0$.

ii) If $m\leq N$, then 

\begin{equation}
 p_m^{MAX} \leq min_{1 \leq r \leq m}  ~{ m \over r}
( \lambda_{m-r+1} + \lambda_{m-r+2} +
\cdots  + \lambda_N ).
\end{equation}

\subsection{The number of Schmidt decomposition terms
can never increase: Part i) of Theorem 1}

The following Lemma is useful.

{\bf Lemma 1}: The number of terms in a Schmidt decomposition
can {\it never} increase under local measurements and
classical communications\cite{charles}.

{\bf Proof}: Let us suppose that the initial state
$| \Phi \rangle = \sum_{i=1}^N \sqrt{ \lambda_i}
| a_i \rangle |b_i \rangle$
has only
$N$ non-vanishing terms in its Schmidt decomposition.
For each measurement outcome $l$ on $| \Phi \rangle$,
the resulting state $P^{Alice}_l | \Phi \rangle =
 \sum_{i=1}^N \sqrt{ \lambda_i}
| a^l_i\rangle  |b_i \rangle
$ [where $| a^l_i \rangle $ is the projected
state $P^{Alice}_l | a_i\rangle$] can be expressed as a
sum of $N$ terms. Consequently, its
Schmidt decomposition must have at most $N$ terms. QED.

{\bf Proof of Part i) of Theorem~1}: As a corollary of
Lemma~1,
for an initial state
$| \Phi \rangle = \sum_{i=1}^N \sqrt{ \lambda_i}
| a_i \rangle  |b_i \rangle$
with only
$N$ non-vanishing terms in its Schmidt decomposition,
$p^{MAX}_m = 0$, if $ m > N$. QED.

This leads to the following apparent paradox.  Suppose
Alice and Bob share $s$ standard singlets. What is the
probability that they can gamble successfully and get $S $ ($ >s $) singlets?
Naively, one might expect the probability to be non-zero:
One may use quantum data dilution\cite{BBPS} to
dilute
$s$ standard singlets into 
say $S$ pairs of $| \Phi \rangle$ each of entanglement 
$E( | \Phi \rangle) = s/S$ and then
apply the Procrustean (i.e., local
filtering) method\cite{BBPS}
of
entanglement gambling to each of $S$ pairs of $| \Phi \rangle$.
For each $| \Phi \rangle$, the Procrustean method
gives a non-zero
probability say, $p'$, of getting a maximally entangled pair
out of it. So, it looks as if there would be a non-zero probability
$(p')^S$ of getting $S$ singlets from $s$ singlets.
But, as we have seen above
this argument is erroneous - the probability of
getting $S$ singlets out of gambling with $s$ singlets is
strictly zero.
The reason is that quantum data dilution is
an {\it inexact} process which is valid only on average. 

\subsection{An Upper Bound on $p^{MAX}_m$: Part ii) of Theorem~1}

It remains to prove Part ii) of Theorem~1.
It is convenient to introduce the following notation.

{\bf Notation}: ($B^m_r$)
We denote the $r$-th bound
in Theorem 1 by $B^m_r$.
i.e.,
\begin{equation}
B^m_r \equiv
{m \over  r} (\lambda_{m-r+1} + \lambda_{m-r+2} +
\cdots  + \lambda_N ).
\end{equation}

{\bf Restatement of Part ii) of Theorem 1}: Given a state
$| \Psi \rangle $ with the ordered Schmidt decomposition
$| \Psi \rangle = \sum_{i=1}^N \sqrt{ \lambda_i}
| a_i \rangle  |b_i \rangle$,
the
supremum probability
$p^{MAX}_m$ 
of obtaining an m-ME-state out of manipulating $| \Psi \rangle $
satisfies
a set of constraints $  p^{MAX}_m  \leq
B^m_r $
for $ 1 \leq r \leq m$.

{\it Idea of the proof of Part ii) of Theorem 1}: For a fixed $r$, if the
right hand side, $B^m_r$, is zero, then
there are only $ m-r $ terms in
the Schmidt decomposition of $| \Psi \rangle$. From
Lemma~1,
Alice will definitely fail to get an m-dimensional maximally
entangled 
pair
state  because there will be at most
$m-r$ terms in the Schmidt decomposition of the resulting state.
We will turn this argument around to show the
following. If Alice does
succeed,
the remaining $r$ (i.e., from
$m-r + 1$-th to $m$-th) terms in the maximally entangled state must have
come from the remaining (i.e., from
$m-r + 1$-th to $N$-th)
terms of the Schmidt decomposition of the original state $| \Phi \rangle$.
[Surprisingly, classical reasoning
is, in fact, valid here.
This is because when one considers
the reduced density matrix of Alice, Bob's system provides
a ``record'' for its history. Therefore, no interference
effect is possible. See below.]
Let us multiply both sides of the inequality and
consider the following new
inequality:
$  r p^{MAX}_m /m
 \leq rB^m_r /m$.
Now the left hand side of the new
inequality is
simply the probability that Alice's state gets projected into the remaining
$r$ terms. [There is a supremum probability $p^{MAX}_m$ of 
getting successfully an m-dimensional maximally
entangled state and
a conditional probability $r/m$ of
getting projected in an $r$-dimensional subspace of
the $m$-dimensional space in the support of Alice's system.]
It must therefore be constrained by the probability of
Bob's system getting projected
into the space spanned by the $m-r + 1$-th to $N$-th
terms in $| \Phi \rangle$, which is given by the right hand side.

{\bf Proof of Part ii) of Theorem 1}: Given an initial state
$| \Phi \rangle$, for $1 \leq r \leq m$,
we decompose $| \Phi \rangle = | \Phi_1^r \rangle +
| \Phi_2^r \rangle $ where $| \Phi_1^r \rangle =
\sum_{i=1}^{m-r} \sqrt{ \lambda_i}
| a_i \rangle  |b_i \rangle$ [Define $| \Phi_1^m \rangle =0$.] and
$| \Phi_2^r \rangle =
\sum_{i=m-r+1 }^{N} \sqrt{ \lambda_i}
| a_i \rangle  |b_i \rangle$. [Define $| \Phi_2^r \rangle =0$
whenever $N < m-r +1$.]
Alice and Bob now attempt to manipulate
$| \Phi \rangle$ into an m-ME-state.
Alice can divide up the outcomes into two
sets: $\{s_1,s_2,  \cdots, s_p \}$ (success) and 
$\{f_1, f_2, \cdots, f_q \}$ (failure). Let us consider
a {\it successful} outcome $s_l$. Then
$P_{s_l} | \Phi \rangle = P_{s_l} | \Phi_1^r \rangle +  P_{s_l}
| \Phi_2^r \rangle $ is an m-ME-state. Denoting by 
$\rho^{s_l}_A $ (similarly $\rho^{r,s_l}_{A,i} $ where $i=1~{\rm or}~2$)
the {\it un-normalized} density matrix
${\rm Tr}_B P_{s_l} | \Phi \rangle \langle \Phi | P_{s_l}^{\dagger}$
(similarly ${\rm Tr}_B
P_{s_l} | \Phi_i^r \rangle \langle \Phi_i^r | P_{s_l}^{\dagger}$
where $i=1~{\rm or}~2$ respectively), we have
$\rho^{s_l}_A = \rho^{r,s_l}_{A,1} + \rho^{r,s_l}_{A,2}$.

We emphasize that the interference term arising from 
${\rm Tr}_B P_{s_l}|  \Phi_1 \rangle \langle \Phi_2 |P_{s_l}^{\dagger}$
is identically
zero. This is because,
when one considers the reduced density matrix of Alice,
Bob's system provides a ``record'' for its history.
In taking the partial trace over Bob's system, all the interference
terms disappear. It is very interesting that classical intuition
is valid here. This greatly simplifies our discussion.

The supports satisfy $supp (\rho^{r,s_l}_{A,1}) \subset
supp (\rho^{s_l}_A)$. Since $supp (\rho^{r,s_l}_{A,1})$ has dimension
at most $m-r$ and yet $supp (\rho^{s_l}_A)$ has dimension $m$
($P_{s_l} | \Phi \rangle$ is an m-ME-state.), we can
pick $r$
orthonormal vectors $| u^{s_l}_1 \rangle , | u^{s_l}_2 \rangle , \cdots,
| u^{s_l}_r \rangle $ in $supp (\rho^{s_l}_A)$
such that $\langle u^{s_l}_i  | v \rangle =0$
for all $| v \rangle \in supp (\rho^{r,s_l}_{A,1})$.
Let us define the projection operator $P^r_{u^{s_l}}
= \sum_{i=1}^r |u^{s_l}_i \rangle  \langle u^{s_l}_i| $. From
its definition, it is clear that $P^r_{u^{s_l}} \rho^{r,s_l}_{A,1}
P^{\dagger r}_{u^{s_l}} = 0$.
For a fixed but arbitrary strategy of entanglement concentration,
let us denote by $p^{arb}_m$ the probability of successfully getting
an $m$-ME-state.
Therefore,
\begin{eqnarray}
& & r p^{arb}_m /m \nonumber \\
&=&
{\rm Tr}_A \left( \sum_{s_l} P^r_{u^{s_l}} \rho_A^{s_l} 
P^{\dagger r}_{u^{s_l}}
\right) \nonumber \\
&=&
 {\rm Tr}_A  \left( \sum_{s_l} P^r_{u^{s_l}} \rho_{A,1}^{r, s_l} 
P^{\dagger r}_{u^{s_l}}
\right)  + 
{\rm Tr}_A  \left( \sum_{s_l} P^r_{u^{s_l}} \rho_{A,2}^{r, s_l} 
P^{\dagger r }_{u^{s_l}}
\right) 
\nonumber \\
&=&
 {\rm Tr}_A  \left( \sum_{s_l} P^r_{u^{s_l}} \rho_{A,2}^{r, s_l} 
 P^{\dagger r}_{u^{s_l}}
 \right)
 \nonumber \\
&=&
 {\rm Tr}_A {\rm ~Tr}_B
\left( \sum_{s_l} P^r_{u^{s_l}} P_{s_l} | \Phi^r_2 \rangle
 \langle  \Phi^r_2 |   P^{\dagger}_{s_l} P^{\dagger r}_{u^{s_l}}
  \right)
\nonumber \\
&\leq &
 {\rm Tr}_A {\rm ~Tr}_B
| \Phi^r_2 \rangle
\langle  \Phi^r_2 |
\nonumber \\
&=&
 \lambda_{m-r +1} +  \lambda_{m-r +2} + \cdots +  \lambda_{N} \nonumber \\
&=&
 r B^m_r/m
 ,
\label{theo}
\end{eqnarray}
for $1 \leq r \leq m$.
The equality sign in the second line holds because
$\rho_A^{s_l}$ is proportional to the identity matrix in
a $m$-dimensional space and its trace is proportional to its
probability of occurring. Since the total probability of success is
$p^{arb}_m$ and $P^r_{u^{s_l}}$ projects an  $m$-ME-state  into an
$r$-dimensional
subspace of the $m$-dimensional space, the probability of this
occurring is clearly $r p^{arb}_m/ m$.

Now, one takes the supremum over all entanglement
manipulation strategies
in Eq.\ (\ref{theo}) to find
that $  p^{MAX}_m   \leq { m \over r} \left(
\lambda_{m-r+1} + \lambda_{m-r+2} +
\cdots  + \lambda_N \right)
= B^m_r $
for $ 1 \leq r \leq m$. QED.

\section{Optimal Strategy and value of $p^{MAX}_m$: Theorem~2}
\label{s:optimals}
Theorem 1 gives an upper bound to the probability $p^{MAX}_m$.
We now prove that an optimal strategy actually saturates
this bound. In other words, we have:

{\bf Theorem 2}: Given a state
$| \Psi \rangle = \sum_{i=1}^N \sqrt{ \lambda_i}
| a_i \rangle  |b_i \rangle$
(where $\lambda_1 \geq \lambda_2 \geq \cdots \geq \lambda_N $)
with only
$N$ non-vanishing terms in its Schmidt decomposition.
There exists a way to convert $\Psi$ into an m-dimensional maximally entangled 
state  with probability  
 $  {\rm min}_{ r \in \{1,2, \cdots, m \} }
{m \over r} (\lambda_{m-r+1} + \lambda_{m-r+2} +
\cdots  + \lambda_N ) = {\rm min}_r  B^m_r $.

{\bf Proof of Theorem~2}:
Let us separate the proof into two cases: (a) ${\rm min}_r  B^m_r = 1$
and (b) ${\rm min}_r  B^m_r < 1$.

\subsection{Case (a) of Theorem 2}

Case (a): Let ${\rm min}_r  B^m_r = 1$. We shall prove that 
for an optimal strategy, the
probability of getting an $m$-ME-state
is $1$.

It is convenient to start with a simple case, namely transforming maximally 
entangled states into maximally entangled states of lower dimension. 
We will prove the following.

{\bf Lemma~2}:
There is a way of transforming with probability 1 any maximally
entangled state into a maximally entangled state of lower dimension.
Consequently, $p^{MAX}_r  \leq p^{MAX}_s $ if $ r \geq s \geq 1$.

{\bf Proof}:
First, consider the case $r=3$ and $s =2$.
(Here we omit the obvious normalization factors.) A maximally
three-dimensionally entangled
state has the Schmidt decomposition
$| u \rangle_{AB} = | 1\rangle_A  | 1 \rangle_B +
| 2\rangle_A | 2 \rangle_B + | 3 \rangle_A | 3 \rangle_B  $.
We now show that it can be reduced with certainty to a standard singlet
$| 1\rangle_A  | 1 \rangle_B +
| 2\rangle_A | 2 \rangle_B $.
Suppose Alice prepares an ancilla in the state $|0 \rangle_a$ and evolves
the system in such a way that $| 0 \rangle_a | 1 \rangle_A
\to ( | 2 \rangle_a + | 3 \rangle_a )  | 1 \rangle_A $,
$| 0 \rangle_a | 2 \rangle_A
\to ( | 1 \rangle_a + | 3 \rangle_a ) | 2 \rangle_A $,
and
$| 0 \rangle_a | 3 \rangle_A
\to ( | 1 \rangle_a + | 2 \rangle_a ) | 3 \rangle_A $.
The entire state will evolve as follows:

\begin{eqnarray}
& & |0 \rangle_a | u \rangle_{AB} \nonumber \\
&=& |0 \rangle_a ( | 1\rangle_A  | 1 \rangle_B +
| 2\rangle_A | 2 \rangle_B + | 3 \rangle_A | 3 \rangle_B ) \nonumber \\
&\to&  |211\rangle_{aAB} + | 311 \rangle_{aAB} + | 122 \rangle_{aAB}  
  +    |322 \rangle_{aAB}  \nonumber \\
& &  + |133 \rangle_{aAB}+ |233 \rangle_{aAB}  \nonumber \\
&=&   |1 \rangle_a ( | 22 \rangle_{AB} + | 33 \rangle_{AB}   ) +
      |2 \rangle_a ( | 11 \rangle_{AB} + | 33 \rangle_{AB}   ) \nonumber \\
& & + |3 \rangle_a ( | 11 \rangle_{AB} + | 22 \rangle_{AB}   ) .
\end{eqnarray}

Now Alice measures the state of her ancilla and obtains a singlet shared with
Bob. The exact singlet which is obtained depends on the result of Alice's
measurement, but it can always be transformed into the standard one
${  1 \over \sqrt{2} } ( | 11 \rangle_{AB} + | 22 \rangle_{AB} ) $.
This can be realized by Alice communicating to Bob the result of her
measurement, such that both of them know which singlet has been obtained
and then having both of them perform the appropriate unitary rotations.

A similar proof can be constructed to show that,
starting with a $k$-state (a maximally entangled pair of $k$-state particles),
Alice and Bob can with probability $1$ convert it
to a $(k-1)$-state (maximally entangled pair of $(k-1)$-state particles).
See Appendix~\ref{a:prooflemma2} for details.~QED.

We remark that using Lemma~2
one can convert with probability 1 a maximally entangled
state
of dimension $i$ into $r$ standard singlets provided that $i\geq 2^r$. 
Just note that,
as mentioned before, $r$ standard singlets are equivalent to a single
$2^r$-dimensional maximally entangled state, and use the above lemma.  This
simplifies a related discussion made in Ref.\ \cite{BBPS} and raises the
probability of success from about $ 1 - \epsilon$ to $1$.

Now we turn to the general case.
The first thing to notice is
that the condition ${\rm min}_rB^m_r=1$ is completely 
equivalent with the constraint that the largest normalized
Schmidt coefficient is smaller 
or  equal to $1/m$. This is because of the following.
\begin{equation}
\lambda_{m-r} \leq \cdots \leq \lambda_1  \leq 1/m
\end{equation}
implies that
\begin{equation}
\lambda_1 + \lambda_2 + \cdots + \lambda_{m-r} \leq { 1 \over m} (m-r) . 
\end{equation}
Since $\lambda_1 + \lambda_2 + \cdots + \lambda_{N} = 1$,
we find that
\begin{equation}
\lambda_{m-r
+1} + \cdots + \lambda_{N} \geq 1 - { 1 \over m} (m-r) = { r \over m},
\end{equation}
which is equivalent to
\begin{equation}
B^m_r = { m \over r} \left( \lambda_{m-r
+1} + \cdots + \lambda_{N}\right) \geq 1 .
\end{equation}
Recall also that $B^m_m =1$. Therefore,
we conclude that, if $\lambda \leq 1/m$, then ${\rm min}_rB^m_r=1$.
Conversely, if $\lambda_1 >  1/m$, $B^m_{m-1} < 1$.
Combining these two results, we see that
${\rm min}_rB^m_r=1$ iff $\lambda_1 \leq 1/m$.

{\it Idea of the Proof of Case (a) of Theorem~2}: Naively, one might
proceed by extracting a $m$-ME-state from $\Psi$ iteratively.
At each step, the state $\Psi' = \Psi'_1 + \Psi'_2 $ such that
$\Psi'_1$ is an (unnormalized)  $m$-ME-state and
$\Psi'_2 $ is residual state that, when properly
normalized, still satisfies
${\rm min}_r  B^m_r = 1$. One simple way
to ensure that ${\rm min}_r  B^m_r = 1$ (or $\lambda_1 \leq 1/m$) is
always satisfied by $\Psi'_2 $ (if properly normalized) is to
allow {\it only} the first $m$ Schmidt terms to contribute to $\Psi'_1$
and, therefore, the $\lambda_1$ term of $\Psi'_2 $ decreases fast
enough.

However, this does not quite work as an iterative procedure.
The reason is that, at some point of such a procedure,
the $m$-th Schmidt coefficient of the state $\Psi$
will become {\it degenerate} with the $m+1$-th and possibly
other coefficients. In other words,
$\lambda_m =\lambda_{m+1}$, etc.
Dealing with this problem is one of the major
technicalities in the proof.
Let us start by making the following definition.

{\bf Definition~5}: (precursor state) Consider a state of the form
\begin{equation}
|\Psi_{\rm pre}^{m,p,q} \rangle = { 1 \over \sqrt{m} } \left( \
\sum_{j=1}^{m-p} | j \rangle | j \rangle + \sum_{j= m-p+1}^{m +q} ({ p \over
p+ q})^{1/2} | j \rangle | j \rangle  \right)
\label{precursor}
\end{equation}
where $ p > 0$ and $ q \geq 0$. Let us call it a precursor
state of an
$m$-ME-state.

{\it Remark}: Note that the case $q=0$ corresponds to an $m$-ME-state.
For $q > 0$, a precursor is a coherent sum of an $(m-p)$-ME-state
and an $(p+q)$-ME-state. The factor $({ p \over
p+ q})^{1/2}$ in the definition of $|\Psi_{\rm pre}^{m,p,q} \rangle$
is needed for the following important result.

{\bf Lemma~3}: A precursor state of an
$m$-ME-state can be converted with
certainty an $m$-ME-state.

{\bf Proof of Lemma~3}: The proof is essentially a generalization of
the proof of Lemma~2.
See Appendix~\ref{a:prooflemma3}.

In our proof, it is convenient to make use of the
following definition.

{\bf Definition~6}: (m-th Schmidt degeneracy number)
For any pure bipartite state $\Psi$ in
an ordered Schmidt decomposition
$
| \Psi \rangle =
\sum_{i=1}^N \sqrt{\lambda_i}|i\rangle_A
|i\rangle_B$,
let us
define the $m$-th  ($m <N$) Schmidt degeneracy number (or
simply the degeneracy number when there is no
ambiguity) to be the number of Schmidt coefficients
that are degenerate with $\lambda_m$.

{\bf Proof of Case a) of Theorem~2}:
Consider the entanglement manipulation of a general state
$| \Psi \rangle
= \sum_{i=1}^N \sqrt{\lambda_i} | i \rangle | i \rangle $
satisfying ${\rm min}_r B^m_r =1$.
We construct a multi-step procedure such that in each step Alice and 
Bob either: 

i) obtain a precursor state which, as shown in Lemma~3,
can readily be reduced with probability $1$ to
an m-dimensional maximally entangled state; or 

ii) obtain a residual state whose (m-th) Schmidt
degeneracy number is increased by $1$,
while still obeying the relation  ${\rm min}_r  B^m_r = 1$ when
properly normalized. 

If Alice and Bob obtain an m-ME-state,
they have accomplished their task. If they 
get a residual state, they repeat 
the procedure.  Since with each step the residual state 
increases its degeneracy number by $1$,  we are certain that in a 
finite number of steps ($\leq N$) either Alice and Bob obtain an m-ME-state,
or end up
with a residual state $\Phi_N$,
which, by Lemma~2,
can subsequently be converted with certainty to an
$\Phi_m$.

We now describe each step in more detail.
%
%For simplicity, we divide each step into two stages. In the first stage Alice 
%and Bob either obtain the appropriate rezidual state, or a ``precursor" of an 
%m-state. In the
%second stage, the ``precursor" is converted  with certainty into 
%an m-state.
%
%Explicitely, the procedure is the following. 
%
%Let the initial state of the pair plus Alices ancilla be 
%
%$$|0\rangle_a|\Psi\rangle=|0\rangle_a\sum_{i=1}^N\sqrt{\lambda_i}|i\rangle_A
%|i\rangle_B$$
Suppose the initial state in ordered Schmidt decomposition is
\begin{equation}
| \Psi \rangle =
\sum_{i=1}^N \sqrt{\lambda_i}|i\rangle_A
|i\rangle_B .
\end{equation}
Suppose further
that $\lambda_m$ 
is $(p+q)$-fold degenerate such that
\begin{equation}
\lambda_{m-p+1}=...=\lambda_m=...=\lambda_{m+q}.
\end{equation}

The decomposition of $| \Psi \rangle$ into a precursor and
a residual state is done by the attachment of
an ancilla prepared in the state
$| 0 \rangle_a $ and a subsequent measurement by Alice.
For $1 \leq i \leq m-p$, the evolution goes as:
\begin{eqnarray}
&    & \sqrt{\lambda_i}   | 0 \rangle_a |i\rangle_A  \nonumber \\
& \to& \sqrt{ a \over m}  | 1 \rangle_a | i \rangle_A 
+  \sqrt{  \lambda_i - {a \over m} } | 0 \rangle_a | i \rangle_A ,
\end{eqnarray}
where $| 0 \rangle_a $ and $| 1 \rangle_a$ are orthonormal.
For $m-p +1 \leq i \leq m+q$, it goes as:
\begin{eqnarray}
 &   & \sqrt{\lambda_i} | 0 \rangle_a |i\rangle_A  \nonumber \\
 & \to& \sqrt{ ({a \over m})({p \over p+q}) } | 1 \rangle_a | i \rangle_A
+  \sqrt{  \lambda_i - ({a \over m}) ({p \over p+q})}
 | 0 \rangle_a | i \rangle_A .
\end{eqnarray}
For $m+q +1 \leq i \leq N$, the state is unchanged, i.e.,
\begin{equation}
| 0 \rangle_a |i\rangle_A \to | 0 \rangle_a |i\rangle_A .
\end{equation}

Hence, we find that
\begin{eqnarray}
&     & | 0 \rangle_a | \Psi \rangle \nonumber \\
& \to &    \sqrt{a} | 1 \rangle_a |\Psi_{\rm pre}^{m,p,q} \rangle +
 \sqrt{ 1 -a } | 0 \rangle_a | \Psi_{\rm res} \rangle
\end{eqnarray}
where
\begin{equation}
|\Psi_{\rm pre}^{m,p,q} \rangle = 
{ 1 \over \sqrt{m} } \left( \
\sum_{i=1}^{m-p} | i \rangle | i \rangle
+ \sum_{i= m-p+1}^{m +q} ({ p \over
p+ q})^{1/2} | i \rangle | i \rangle  \right)
\end{equation}
is the precursor and
\begin{eqnarray}
| \Psi_{\rm res} \rangle  &=& ( 1 -a )^{- 1/2}
 \left[ \sum_{i = 1}^{m-p} \sqrt{ \lambda_ i - { a \over m} } | i \rangle
 |i \rangle \right. \nonumber \\
& & + 
\sum_{i = m-p+1 }^{m+q} \sqrt{ \lambda_ i - ({ a \over m})
({ p \over p+q}) }  | i \rangle |i \rangle \nonumber \\
& & + \left.
\sum_{i= m +q +1}^N \sqrt{ \lambda_ i} | i \rangle |i \rangle \right]
\end{eqnarray}
is the
residual state and $a$ is the minimal
value needed for a new degeneracy to occur in Schmidt coefficients of
the residual state $| \Psi_{\rm res} \rangle$. i.e., $a = {\rm min}
\left( {m ( p+q) \over q } ( \lambda_{m-p}  - \lambda_{m-p+1 }) ,
{m ( p+q) \over p } ( \lambda_{m+q}  - \lambda_{m+q+1 }) \right) $,
thus achieving either (1) $\lambda'_{m- p} = \lambda'_{m- p+1 }$
or (2) $\lambda'_{m+q} = \lambda'_{m+q+1 }$.

Now Alice measures the state of the ancilla. If the outcome is ``1'',
she gets a precursor state which can
be converted with certainty to an $m$-ME-state.
If the outcome is ``0'', she gets
a residual state with its degeneracy number increased by $1$.

It is also easy to see that, just like the original state $\Psi$, the 
intermediate residual state $| \Psi_{\rm res} \rangle $
also has the property that ${\rm min}_rB^m_r=1$.
The final residual state will be totally degenerate and,
hence, has the form $\Phi_N$.
This multi-step method establishes our proof. QED.

%After this first stage Alice measures the state of her ancilla.
%If she finds the ancilla in the state $|0\rangle_a$, that is,
%she obtains the residual state, she 
%repeats the above procedure. If she obtains the ``precursor" state, (i.e. she 
%finds the ancilla in the state $|0\rangle_a$) she converts it into an 
%m-dimensional maximally entangled state by the following actions, similar to 
%those employed in Section (??)
%
%$$|0\rangle_a| i \rangle_A\rightarrow \bigl({1\over{\sqrt 
%{p+q+1}}}\sum_{j=1}^{p+q+1} |j\rangle_a\bigr)| i \rangle_A,~~~for~1\leq j\leq 
%m-p-1$$
%
%$$|0\rangle_a| i \rangle_A\rightarrow \bigl({1\over{\sqrt{p+q}}}\sum_{j=1; 
%j\n%eq 
%i}^{p+q} |j\rangle_a\bigr)| i \rangle_A,~m-p\le j\leq m+q.$$
%
%Upon measuring the state of the ancilla, Alice and Bob end in a new  
%``precursor" type state, closer to the
%desired m-state. Repeating the procedure 
%$q$ times the original precursor () is completely transformed into an m-state.

\subsection{Properties of $B^m_r$}
\subsubsection{Lemma~4}

Before moving to Case (b), let us prove some lemmas.
For any initial state $|  \Psi \rangle $, the bounds in
theorem 1, $B^m_r= {m \over r} ( \lambda_{m-r+1 } +  \lambda_{m-r+2 }
+ \cdots +
\lambda_N )$, obey the following.

{\bf Lemma~4}:: If $B^m_{r+1} > B^m_r$, then $B^m_{r+2} > B^m_{r+1}$.

{\it Remark}: In other words, for a fixed $m$, consider $B^m_r$ as a function
of $r$. Once it starts to increase, it will continue to do so.

{\bf Proof}: See Appendix~\ref{a:prooflemma4}.

\subsubsection{Lemma 5}

By adding the condition (which is valid for case (b) of Theorem~2)
that  ${\rm min}_r  B^m_r < 1$, the following Lemma can be
proven.

{\bf Lemma~5}: Given  ${\rm min}_r  B^m_r < 1$, there exists a {\it unique}
$r_1$ such that $B^m_1 \geq B^m_2 \geq  \cdots \geq B^m_{r_1} <
B^m_{r_1 +1 } < \cdots < B^m_m =1$.

{\bf Proof}: See Appendix~\ref{a:prooflemma5}

Remark: Since $B^m_r$ is defined to be $
{m \over r} ( \lambda_{m-r+1 } +  \lambda_{m-r+2 }
+ \cdots +
\lambda_N )$, in terms of $\lambda_i$'s, the conditions
that
$B^m_1 \geq B^m_2 \geq  \cdots \geq B^m_{r_1} <
B^m_{r_1 +1 } < \cdots < B^m_m =1$ can be written as
the following set of equations:
\begin{eqnarray}
\lambda_{m-1 }& \leq& \lambda_{m } + \lambda_{m+1 } + \cdots + \lambda_{N}
\nonumber \\
\lambda_{m-2 }& \leq&  { 1 \over 2}
( \lambda_{m-1 } + \lambda_{m } + \cdots + \lambda_{N})
\nonumber \\
\cdots & \leq& \cdots \nonumber \\
\lambda_{m-r_1 +1}& \leq&  { 1 \over (r_1 -1 ) }
( \lambda_{m-r_1+2  } +\lambda_{m-r_1 +3  } + \cdots + \lambda_{N})
\nonumber \\
\lambda_{m-r_1 }& > &  { 1 \over (r_1 ) }
( \lambda_{m-r_1 +1  } +\lambda_{m-r_1 +2  } + \cdots + \lambda_{N})
\nonumber \\
\cdots & > & \cdots \nonumber \\
\lambda_1 & > &  { 1 \over (m-1) }
( \lambda_{2 } +\lambda_{3  } + \cdots + \lambda_{N}) .
\label{yan}
\end{eqnarray}

Inspired by the above discussion,
let us consider the following set of equations.
\begin{eqnarray}
\lambda_{m-1 }& \leq& \lambda_{m } + \lambda_{m+1 } + \cdots + \lambda_{N}
\nonumber \\
\lambda_{m-2 }& \leq&  { 1 \over 2}
( \lambda_{m-1 } + \lambda_{m } + \cdots + \lambda_{N})
\nonumber \\
\cdots & \leq& \cdots \nonumber \\
\lambda_{m-r}& \leq&  { 1 \over r }
( \lambda_{m-r+1  } + \lambda_{m-r +2  } + \cdots + \lambda_{N})
\nonumber \\
\cdots & \leq & \cdots \nonumber \\
\lambda_1 & \leq &  { 1 \over (m-1) }
( \lambda_{2 } +\lambda_{3  } + \cdots + \lambda_{N}) .
\label{constraints}
\end{eqnarray}
Consider putting $\lambda_{m-1}, \lambda_{m-2}, \cdots, \lambda_1$
into the left hand side of Eqs.~(\ref{constraints}) one by one,
we find from Eqs.~(\ref{yan})
that $\lambda_{m-1}, \lambda_{m-2} , \cdots,  \lambda_{m-r_1+1  }$
satisfy Eqs.~(\ref{constraints}) whereas $\lambda_{m-r_1}, \lambda_{m-r_1-1},
\cdots,   \lambda_1$ violate Eqs.~(\ref{constraints}).
Let us focus on the point of first violation, namely $\lambda_{m-r_1}$.
We notice that the maximal value of $\lambda^{\rm max}_{m-r_1}$ that
will still satisfy  Eq.~(\ref{constraints}) is
\begin{eqnarray}
\lambda^{\rm max}_{m -r_1}
&\equiv& { 1 \over r_1} ( \lambda_{m - r_1 +1}
+ \cdots + \lambda_N ) \nonumber \\
    ~   & = & { B^m_{r_1} \over m} \nonumber \\
     ~  & = & {\rm min}_r  B^m_{r}.
\label{lammax}
\end{eqnarray}

With lemmas 4 and 5 proven, we now return to the proof of case (b) of
theorem 2.

\subsection{Case (b) of theorem 2}

Case (b): ${\rm min}_r  B^m_r < 1$.

{\it Idea of our proof}:
We construct an explicit strategy which saturates
the bound $p_m = {\rm min}_r B^m_r$ as follows.
By attaching an ancilla prepared in the state $| 0 \rangle_a$
to the system $| \Psi \rangle$, Alice
divides up $| \Psi \rangle $ into two
pieces---successful and failing pieces---by the following evolution:
\begin{equation}
| 0 \rangle_a | \Psi \rangle = | 1 \rangle_a | \Psi_s \rangle +
| 0 \rangle_a|  \Psi_f \rangle
\label{sf}
\end{equation}
where $| 0 \rangle_a$ and $ | 1 \rangle_a $ are orthonormal states
of the ancilla,
$| \Psi_s \rangle$ (when properly normalized
belongs to case (a), i.e., ${\rm min}_r  B^m_r = 1$
and hence) gives a probability $1$ of success
and $|  \Psi_f \rangle$ (has less than $m$ terms in its Schmidt decomposition
and hence) gives a probability $0$ of success.
Alice now reads off the state of the ancilla.
A state $| 1 \rangle_a $ indicates a success and
$| 0 \rangle_a$ a failure.
One can then read off the probability of success
of this explicit strategy from the norm of $| \Psi_s \rangle$.
It turns out to be equal to ${\rm min}_r B^m_r $.

{\bf Proof of case (b) of Theorem 2}:
Recall from Eq.~(\ref{lammax})
that the maximal acceptable value of the $(m-r_1)$-th 
Schmidt coefficient for it to satisfy Eq.~(\ref{constraints}) is
\begin{equation}
\lambda^{\rm max}_{m -r_1}= { B^m_{r_1} \over m}={\rm min}_r  B^m_{r} .
\end{equation}
Now the successful piece $| \Psi_s \rangle$ in Eq.\ (\ref{sf}) is obtained
by trimming the redundant contribution to
$ \lambda_1, \lambda_2, \cdots, \lambda_{m -r_1}$ from $ | \Psi \rangle $.
This is done by the attachment of an ancilla
prepared in the state $| 0 \rangle_a$. The evolution goes as follows:
\begin{eqnarray}
 \sqrt{\lambda_i } | 0 \rangle_a | i \rangle_A 
 & \to & \sqrt{\lambda^{\rm max}_{ m - r_1} } | 1 \rangle_a | i \rangle_A 
\nonumber \\
 &  & + \sqrt{\lambda_i  -  \lambda^{\rm max}_{ m - r_1} } | 0 \rangle_a
| i \rangle_A 
\end{eqnarray}
for $ 1 \leq i \leq m -r_1$.
For $ m - r_1 +1 \leq i \leq N $, the evolution is
\begin{equation}
 \sqrt{\lambda_i } | 0 \rangle_a | i \rangle_A 
 \to \sqrt{\lambda_i } | 1 \rangle_a | i \rangle_A .
\end{equation}

Alice now reads off the state of her ancilla.
We shall argue in the following
paragraph that an outcome ``0'' means that
Alice has failed in getting an $m$-ME-state whereas
an outcome ``1'' means that she has succeeded in
obtaining a state satisfying ${\rm min}_r B_r^m =1$, which by
Sec.~VI~B can be
reduced with certainty to an $m$-ME-state.

If the outcome is ``0'', the resulting
(failing) state $|  \Psi_f \rangle$ has unnormalized
Schmidt coefficients $\lambda_1 - \lambda^{\rm max}_{m -r_1}, 
\lambda_2 - \lambda^{\rm max}_{m -r_1}, \cdots, \lambda_{m-r_1}
 - \lambda^{\rm max}_{m -r_1}, 0, \cdots, 0$. Since it has at most
$m-r_1$ terms in its Schmidt decomposition, it follows from
Lemma~1
that it gives a zero probability of getting a $m$-ME-state.
On the other hand, if the outcome is ``1'',
the  {\it un-normalized} Schmidt coefficients of
the resulting (successful) state $ | \Psi_s \rangle $
are given by $ \lambda^{\rm max}_{m -r_1}, \cdots, \lambda^{\rm max}_{m -r_1},
\lambda_{m - r_1 +1}, \lambda_{m - r_1 +2}, \cdots ,\lambda_N $.
i.e., the first $m -r_1$-th Schmidt coefficients are all replaced by
$ \lambda^{\rm max}_{m -r_1}$.
By construction $|  \Psi_s \rangle$
belongs to Case (a) of Theorem 2. Therefore, it always succeeds to
give an $m$-ME-state. Moreover, using Eq.~(\ref{lammax})  it
has a norm

\begin{eqnarray}
& & (m - r_1) \lambda^{\rm max}_{m -r_1} + \lambda_{ m - r_1 + 1} + \cdots +
\lambda_{N} \nonumber \\
& =& { m \over r_1 } ( \lambda_{m - r_1 +1}
+  \lambda_{m - r_1 +2} + \cdots + \lambda_N )  \nonumber \\
&=& B^m_{r_1}  \nonumber \\
&=& {\rm min}_r B^m_r . 
\end{eqnarray}
This proves that our explicit strategy saturates
the bound and completes our proof for the case (b) of Theorem 2. QED.

Recall that Theorems~1 and 2 combined together are equivalent to
our Main Theorem (Theorem~0). Since we have by now proven
both Theorems~1 and 2, our Main Theorem has been established.

\section{The law of large numbers}
In this section, we derive some constraint on the probabilities of
having large deviations from the average properties.
Consider the question raised in the abstract and the introduction:  Can collective
measurements defeat the law of large numbers?  We now show that the answer is
no.  That is, suppose Alice and Bob share $n$ pairs of particles, each pair in a
state $| \Psi\rangle$ with an entropy of entanglement $E(| \Psi\rangle )$.  We
shall show
in Theorem 3 below that the maximal probability of obtaining $nK$ 
singlets, with $K >
E(| \Psi\rangle )$, goes to zero as $n$ goes to infinity.

Once again, we want to emphasize that this result {\it does not} follow
automatically from the fact that {\it on average} we cannot obtain more than
$nE$ singlets.  Indeed, an average of $nE$ singlets could conceivably be
obtained if with a {\it non-negligible} probability $p=E/K$ we get $nK$ singlets
while with probability $1-E/K$ we get no singlets at all.

{\bf Theorem 3}: In the entanglement manipulation of $n$ pairs $\Psi$,
the optimal probability (over all possible strategies)
of getting $nK$ singlets, $p^{MAX}_{2^{nK}}$, tends
to $1$ ($0$ respectively) when $K < E (| \Psi\rangle)$ ($K >E (|
\Psi\rangle)$ respectively) in the limit $n \to \infty$. 

{\it Remark}: It can also
be shown that, as a function of $K$,
the jump from $0$ to $1$ in the value of $p^{MAX}_{2^{nK}}$
occurs in a region of width $O (n^{-1/2})$ around $E (| \Psi\rangle)$.
We shall skip the proof here.

{\bf Proof of Theorem 3}: That $p^{MAX}_{2^{nK}}$ tends to $1$ in the
large $n$ limit when $ K <E (|
\Psi\rangle)$ follows trivially from
Bennett {\it et al.}'s reversible strategy\cite{BBPS} and
Lemma~2.
Let us now consider the case $K >E (|
\Psi\rangle)$.
Here we view the $n$ pairs $\Psi$ as a
single pair in state ${\tilde \Psi} = \Psi^n $, by
considering all $n$ Alice's (Bob's)
particles to form a single (more complex) quantum system. 
Similarly, the final
$nK$ singlet pairs can be viewed as a single pair in a $2^{nK}$-dimensionally
maximally entangled state.  Then the problem of extracting $nK$ singlets
from
the $n$ pairs $\Psi$ can be rephrased as the problem of extracting an
$2^{nK}$-dimensionally maximally entangled state from $\tilde \Psi$.
The maximal
probability for success is $p^{MAX}_{2^{nK}}$ which can be
bounded by using
Theorem 1.

Let $\tilde\lambda_i$'s represent the Schmidt
coefficients of $\tilde\Psi$; they 
are also the eigenvalues of Alice's reduced density matrix. 
Since Alice's reduced density matrix 
has a product form, (originating from the $n$ pairs $| \Psi\rangle$) its
weight 
must be concentrated on a `typical' space
of dimension roughly $2^{ n E}$. [Here we simply our notation
and use $E $ to denote $E(| \Psi\rangle)$. This is essentially the law
of large numbers in classical probability
theory. See also quantum noiseless source coding
theorem\cite{Sch}.]
Let us pick a $K_0$ such that $K > K_0 > E$. 
Since $K_0 > E$, given any $\delta > 0$,
for a sufficiently large $n$, we have that
$\sum_{i = 2^{nK_0 } }^{t^n} \tilde\lambda_i < \delta $ where $t$ is
the number of terms in the Schmidt decomposition of $| \Psi \rangle$.
[An `atypical' space has a small weight.]
Let us apply theorem 1 to the case $N = t^n$, $m =2^{nK } $ 
and $m -r + 1 = 2^{nK_0 }$. Notice that $r/m > 1/2$ for a sufficiently
large $n$. Hence, $p^{MAX}_m/2 < rp^{MAX}_m /m \leq \sum_{i = m -r + 1}^{t^n}
\tilde\lambda_i < \delta$. Substituting $m =2^{nK } $ back, we get
$p^{MAX}_{2^{n K}} \to 0$ as $n \to \infty$. QED.

In fact, any particular strategy which transforms $n$
copies of the state
$\Psi$ into an average of $nE$ singlets gives a singlet
number probability distribution similar to that of
reversible strategy\cite{BBPS}. This follows immediately from the
result in the next section.

\section{Special Strategies}
\label{s:special}
In the previous sections we were interested in the question of what is the 
maximal probability to transform an arbitrary entangled state $\Psi$ into a 
given maximally entangled state, say $\Phi_m$ (where $m$ is some given fixed 
dimension). What happens to the original state $\Psi$ in those cases in which 
the transformation into $\Phi_m$ is {\it not} successful was not important for 
us. We will now consider special manipulation strategies which are such that
for 
{\it every} outcome the initial state is transformed into some maximally 
entangled state. 
(Note that, by extension of language, we denote direct product states as 
``maximally entangled states of dimension zero"). Such a strategy
${\cal S}$ can 
be characterized by the probabilities $p_0(\cal S)$, $p_1(\cal S)$, ... 
with which the initial state $\Psi$ is transformed into $\Phi_0$,
$\Phi_1$,... 
respectively.

A convenient way to describe this probability distribution is to use
instead of
the probabilities $p_m({\cal S})$ the ``cumulative probability"
$p_m^{tot}
({\cal S})$,

\begin{equation}
p_m^{tot}({\cal S})=\sum_{k\geq m}p_k({\cal S}) .
\label{eq6}
\end{equation}

In the present section we find an upper bound on the cumulative property for 
an arbitrary strategy ${\cal S}$.
\begin{equation}
p_m^{tot}({\cal S})\leq p_m^{MAX}
\label{eq7}
\end{equation}
where $p_m^{MAX}$ is the supremum
probability over all possible strategies to 
convert $\Psi$ into an m-dimensional maximally entangled state
(an $m$-state).
Since $p_m({\cal S})$ represents the probability to convert $\Psi$ into an 
m-state 
by using the particular strategy ${\cal S}$ while $p_m^{MAX}$ represents the 
supremum 
probability (over all possible strategies) to convert $\Psi$ into an m-state,
it is obvious that $ p_m({\cal S})\leq p_m^{MAX}$.
But why should the sum $p_m({\cal S})+p_{m+1}({\cal S})+...$ be smaller than 
$p_m^{MAX}$? 

The reason is that, as we have shown Lemma~2, a
maximally entangled state of dimension 
$k$ can always be converted {\it with certainty}, into a maximally entangled 
state of smaller dimension $m$ ($m<k$). Then, suppose that Alice and Bob, by 
using the strategy ${\cal S}$ convert $\Psi$ into a maximally entangled
state of  dimension $k$
larger than $m$. They can then convert, with certainty, this state 
into a maximally entangled state of dimension equal to $m$. Consequently, by 
appending this reduction strategy to the strategy ${\cal S}$, we
obtain a new strategy ${\cal S'}$ which converts $\Psi$ into an
$m$-state with  
probability $p_m({\cal S'})=
\sum_{k \geq m} p_m({\cal S}) =p^{tot}_m({\cal S})$, (while having zero 
probability 
to convert $\Psi$ into maximally entangled states of dimension
larger than $m$).
Now, as $p_m^{MAX}$ is the supremum
probability (over all possible strategies) of 
converting $\Psi$ into an m-state,
we must have in particular $p^{MAX}_m \geq p_m({\cal S'})=p^{tot}_m({\cal S})$ 
which 
proves the bound in Eq.\  (\ref{eq7}).

\section{Non-existence of Universal Strategy}

As shown in Section \ref{s:special},
for any strategy $\cal S$ which transforms an
arbitrary 
state $\Psi$ into different maximally entangled states $\Phi_m$, the
cumulative 
probability $p_m^{tot}$ of obtaining some maximally entangled state of
dimension $m$ 
or larger is bounded by 
\begin{equation}
p^{tot}_m\leq p_m^{MAX}.
\label{nmax}
\end{equation}
We have also seen in the previous section that for any particular $m$ there 
exists a strategy which saturates this bound (the strategy which yields $\Phi_m$ 
with probability equal to $p_m^{MAX}$ and $\Phi_k$, $k>m$ with zero 
probability). The question is whether there exists a ``universal" strategy 
${\cal S}^{univ}$ whose cumulative distribution saturates this bound for {\it 
all} $m$'s.  The reason we call such a strategy ``universal" is that such a 
strategy, followed by the reduction of some of the final maximally entangled 
states into maximally entangled states of lower dimension could generate any 
possible distribution consistent with the bound (\ref{nmax}).
We shall show however that 
such a universal strategy does not exist.

Proof: We show that
a universal strategy generally
cannot exist for the case $N=3$ and $m=2 $ or $3$.
Consider
\begin{equation}
| \Psi \rangle  = \sqrt{\lambda_1} | 11 \rangle +
\sqrt{\lambda_2} | 22 \rangle + \sqrt{\lambda_3} | 33 \rangle
\end{equation}
with $p^{MAX}_2 = 1$ and $\lambda_2 + \lambda_3 - \lambda_1 \geq 0$.
Assume, by means of contradiction, that a universal strategy does
exist. We shall use projection operators rather than
positive operator valued measures (POVMs)
in our discussion. As noted in Sec. 2, there is no loss of
generality.
Let $P_1, P_2, \cdots, P_r$ be the set of all projection operators
by Alice that give some $3$-state in a particular universal
entanglement manipulation strategy.
By definition, $(P_1 + P_2 + \cdots 
+ P_r) | \Psi \rangle $ has a norm $p^{MAX}_3$.
Note that it follows from Theorem~2 that $p^{MAX}_3 = 3 \lambda_3$.
Since $p^{MAX}_2=1$, it is necessary for a universal strategy
that the residual state
$  | \Psi_r \rangle = ( 1- P_1 -P_2 - \cdots - P_r)
| \Psi \rangle$ has $p^{MAX}_2 =1$.
But this requires the squared eigenvalues of the
reduced density matrix of $  | \Psi_r \rangle$
to satisfy the constraint $\lambda'_2 + \lambda'_3 - \lambda'_1 \geq 0$.
We shall show that this is generally impossible.
The point of our argument
is that, as shown by Lemma~6 below,
the extraction of a $3$-state will lead to
an equal decrease in all three squared eigenvalues (of the reduced density
matrix of $ | \Psi_r \rangle$). i.e., $\lambda'_i = \lambda_i - p^{MAX}_3/ 3
=  \lambda_i- \lambda_3$. Therefore, unless $\lambda_1 = \lambda_2$,
the residual state
$ | \Psi_r \rangle $ has $\lambda'_2 + \lambda'_3 - \lambda'_1 =
\lambda_2 - \lambda_1 <0$,
thus contradicting the requirement that $p^{MAX}_2 (| \Psi_r \rangle) =1$.

In the above proof, we have used the following Lemma.

{\bf Lemma 6}: Consider a state
\begin{equation}
| \Psi \rangle  = \sqrt{\lambda_1} | 11 \rangle +
\sqrt{\lambda_2} | 22 \rangle + \sqrt{\lambda_3} | 33 \rangle
\end{equation}
in Schmidt decomposition. Any strategy that extracts
a 3-ME-state with a probability $p$ from $\Psi$ will
lead to an equal decrease in all three eigenvalues of
the reduced density matrix of the {\it un-normalized} residual state. i.e.
$\lambda'_i = \lambda_i - p/3$ where the
$\lambda'_i$'s are eigenvalues of the reduced
density matrix of the un-normalized residual state.

{\bf Proof of Lemma 6}: See Appendix~\ref{a:prooflemma6}.

\section{Mixed States}  

Let us now consider the case when Alice and Bob share
a mixed initial state $\rho_{\rm ini}$. Since
$\rho_{\rm ini}$ is impure, one generally cannot write it
in terms of Schmidt decomposition. More importantly, even if
$\rho_{\rm ini}$ {\it happens} to be symmetric
under the interchange of Alice and Bob, there is no guarantee that
the intermediate states that they get during the entanglement
manipulation process will
respect such a symmetry\cite{spe}. Therefore,
the symmetry argument
much emphasized in the earlier part of this paper
will no longer be valid.
Manipulations of a mixed state using
two-way communications are generally more advantageous than a one-way
strategy. Indeed, Bennett {\it et al.} have shown that
one-way capacity and two-way capacity for purification
are provably different\cite{BDSW}.

We also proved in Section~\ref{reduction} that,
for a pure bipartite state, entanglement manipulation
strategies with
one-way communications are provably
better than no communications. Notice that
one-way communications is useful for an entanglement
manipulation strategy that has a probability of success
strictly between $0$ and $1$,
but not
for (deterministic) quantum error correction\cite{BDSW}.
The role of communications in entanglement manipulations
deserves future
investigations.

For a mixed state, there are generally four
distinct supremum probabilities to consider:
$p_m^2$, $p_m^{A \to B}$, $p_m^{B \to A}$ and $p_m^0$ corresponding to
entanglement manipulation schemes with two-way communications,
one-way communications
from Alice to Bob, one-way communications from Bob to Alice
and no communications respectively.
While simple bounds on the success probability for manipulating
mixed states may be derived, many interesting questions
remain unanswered. For example, we do not know
the value of $p_{2^{nA}}$ in the asymptotic limit $n \to \infty$
in the region
$ D_0 (\rho) \leq A \leq \ E(\rho)$ where $ D_0 (\rho)$ is the
entanglement of distillation (without any classical communications between
Alice and Bob).

To conclude, we expect the subtle interplay of the concepts
of probability, classical communications, collective
manipulations
and symmetry in the case of mixed states
to be even more challenging than the
pure state case considered in this paper.

\section{open questions on pure states}
Even for the case of a pure initial state, many interesting
questions remain unsolved. For instance, what is
the supremum probability $p_m^0$ of getting an m-ME-state
without any classical communications?
Notice that Bennett {\it et al.}'s reversible
strategy\cite{BBPS} (but not the local filtering strategy\cite{BBPS})
is an example of a strategy which does not
require any classical communications.
It is an open question whether one
can do better than Bennett {\it et al.}'s strategy without
any classical communications.

We emphasize that the symmetry that we have
found here applies not only to entanglement concentration,
but also to all types of entanglement manipulations including
entanglement dilution\cite{BBPS}.
For instance,
the usual procedure of entanglement dilution via teleportation
falls inside our general framework of using a single generalized
measurement by Alice followed by one-way communications of its
result to Bob and a subsequent unitary transformation by Bob.
A more systematic investigation of our formalism in applications
beside entanglement concentration
may prove rewarding.

\section{Acknowledgments}

H.-K. Lo particularly
thanks P. Shor for enlightening discussions which indirectly inspired this
line of research. Our proof of Theorem 1 has been simplified following
a critical comment by R. Jozsa.
S. Popescu thanks C. H. Bennett and J. Smolin
for helpful communications
on their independent proof of Lemma~1.
Useful discussions with R. Cleve, D. Gottesman, D. Leung, M. A. Nielson
and J. Preskill
are greatly appreciated.
Many helpful comments and suggestions from an anonymous referee are
also gratefully acknowledged.
Part of the writing of an earlier version
of this paper was done
during a visit of H.-K. Lo to Quantum Information and Computing
(QUIC) Institute at Caltech, whose hospitality is gratefully acknowledged.

\appendix
\section{Proof of Proposition~1}
\label{a:proofprop1}
Let us write $\Psi$ in its Schmidt decomposition:
\begin{equation}
| \Psi \rangle = \sum_k \sqrt{\lambda_k} | a_k \rangle | b_k \rangle.
\end{equation}
Consider any of Bob's projection operator
\begin{equation}
P^{\rm Bob}_l = \sum_{i,j} m^l_{ij} |b_i \rangle \langle b_j | .
\end{equation}
After the projection, the state he shared with Alice becomes
\begin{eqnarray}
| \Psi^B \rangle &=&
\left( I \otimes P^{\rm Bob}_l \right)| \Psi \rangle \nonumber \\
 &=& \sum_{i,k} \sqrt{\lambda_k} m^l_{ik} |a_k\rangle | b_i \rangle .
\label{eqnA}
\end{eqnarray}
On the other hand, if, instead of Bob, Alice performs a
measurement
using the corresponding operator defined by
\begin{equation}
P^{\rm Alice}_l = \sum_{i,j} m^l_{ij} |a_i \rangle \langle a_j | ,
\end{equation}
an outcome $l$ will give the state
\begin{eqnarray}
| \Psi^A \rangle &=&
\left( P^{\rm Alice }_l  \otimes I \right) | \Psi \rangle \nonumber \\
 &=& \sum_{i,k} \sqrt{\lambda_k} m^l_{ik} |a_i\rangle | b_k \rangle
\label{eqnB}
\end{eqnarray}

Let us consider unitary transformations $U$
($| a_i \rangle \to \sum_p u_{ip} |a_p\rangle $)
and $V$ ($| b_k \rangle \to \sum_q v_{kq} |b_q\rangle $) that
will put $\Psi^A$ in Schmidt decomposition. i.e.,
\begin{equation}
\left( U \otimes V \right) | \Psi^A \rangle = \sum_p \sqrt{\mu_p}
| a_p \rangle | b_p \rangle.
\label{eqn:uv}
\end{equation}
From the definitions of $U$ and $V$ and Eqs.~(\ref{eqnB}) and
(\ref{eqn:uv}), we find that
\begin{equation}
\sum_{ik}  \sqrt{\lambda_k} m^l_{ik} u_{ip}v_{kq} =
\sqrt{\mu_p} \delta_{pq} .
\label{eqn:sumik}
\end{equation}

Now consider $\left( V \otimes U \right) | \Psi^B \rangle $.
\begin{eqnarray}
\left( V \otimes U \right) | \Psi^B \rangle &=&
\sum_{ik} \sum_{pq} \sqrt{\lambda_k} m^l_{ik} v_{kq}u_{ip}
|a_q\rangle | b_p \rangle  \nonumber \\
 &=& \sum_{pq} \sqrt{\mu_p} \delta_{pq} |a_q\rangle | b_p \rangle
\nonumber \\
 &=& \sum_p \sqrt{\mu_p } |a_p\rangle | b_p \rangle ,
\label{eqn:vu}
\end{eqnarray}
where Eq.~(\ref{eqn:sumik}) is used in the second equality.

From Eqs.~(\ref{eqn:uv}) and (\ref{eqn:vu}),
we find that
\begin{eqnarray}
\left( V \otimes U \right) | \Psi^B \rangle &=&
\left( U \otimes V \right) | \Psi^A \rangle \nonumber \\
 | \Psi^B \rangle &=&
\left( V^{-1}U \otimes U^{-1}V \right) | \Psi^A \rangle \nonumber \\
\left( I \otimes P^{\rm Bob}_l \right)| \Psi \rangle
&=& \left( U^A_l \otimes U^B_l \right)
\left( P^{\rm Alice }_l \otimes I 
\right) | \Psi \rangle 
\end{eqnarray}
where $U^A_l = V^{-1}U$ and $ U^B_l= U^{-1}V$.~QED.

\section{Proof of the necessity of one-way communication
in entanglement manipulations of bipartite pure states}
\label{noway}
Definitions~2 and 3 in the main text are needed for this proof.
The basic reason for the necessity of classical
communication is that,
whenever $p_m^{max}$ as defined in the text
is strictly less than 1, Bob generally
needs Alice's help to figure out whether the
entanglement manipulation is
successful or not.

Consider the
example of $| \Psi \rangle = a | 11 \rangle + b | 22 \rangle$
where $a > b > 0$.
We shall first argue that the supremum probability of obtaining
a singlet satisfies
$0 < p^{\rm MAX}_2 < 1$: Since the
local filtering strategy in Ref.\ \cite{BBPS} gives
a non-zero probability of getting a singlet, we
have $p^{\rm MAX}_2 \geq p^{\rm local~
filtering}_2 > 0$.
Moreover, since the entanglement $E( \Psi) < 1$ and
the average entanglement cannot increase upon
entanglement manipulations, the supremum
probability $p^{\rm MAX}_2$ of getting a singlet out of entanglement
manipulations is less than $1$.

Now consider any strategy that gives $0 < p_2 < 1$.
Let us divide up its outcomes into two classes: $\{s_1, s_2, \cdots , s_p \}$
(success) and $\{ f_1 , f_2 , \cdots , f_q \}$ (failure) and
denote the {\it un-normalized} reduced density matrix of Bob for
an outcome $s_i$ ($f_j$) by $ \rho^{Bob}_{s_i}$ ($\rho^{Bob}_{f_j}$).
Since  $0 < p_2 < 1$, Bob needs to determine the outcome of the
entanglement manipulation
by distinguishing with certainty between the two density matrices
$\rho^{Bob}_{success} = \sum_i \rho^{Bob}_{s_i}$
and $\rho^{Bob}_{failure} = \sum_j \rho^{Bob}_{f_j}$.
Now the distinguishability of two density matrices can be
described by
the fidelity\cite{fidelity}
$F({ \rho^{Bob}_{success} \over {\rm Tr}
\rho^{Bob}_{success} }, {\rho^{Bob}_{failure} \over {\rm Tr}
\rho^{Bob}_{failure}} )$.
The detailed definition and properties of the fidelity are
irrelevant for our discussion. It suffices to note the
following fact:
In order to show that it is impossible for Bob to
distinguish with certainty between the two density matrices without
communications from Alice, all we need to prove is that
$F({ \rho^{Bob}_{success} \over {\rm Tr}
\rho^{Bob}_{success} }, {\rho^{Bob}_{failure} \over {\rm Tr}
\rho^{Bob}_{failure}} ) \not= 0$
or equivalently the supports of $\rho^{Bob}_{success}$
and $\rho^{Bob}_{failure}$ are not orthogonal to each other.
The proof of this claim is simple:
Owing to causality, the density matrix of Bob
is conserved throughout Alice's measurement, i.e.,
\begin{eqnarray}
\rho^{Bob}_{success}+ \rho^{Bob}_{failure}
&= & \rho^{Bob}_{initial} \nonumber \\
&= & a^2 |1 \rangle \langle 1 | + b^2  | 2 \rangle \langle 2 |.
\end{eqnarray}
Since $\rho^{Bob}_{initial} $
has a two-dimensional support, $\rho^{Bob}_{success}$
must have a support of at most two dimensions.
On the other hand, as $\rho^{Bob}_{s_i}$ is the reduced
density matrix for a singlet,
$\rho^{Bob}_{success}$, being the
sum of $ \rho^{Bob}_{s_i}$'s, must have a support of at least two dimensions.
Combining these two statements, $\rho^{Bob}_{success}$ has
a support of exactly two dimensions. Now that both
$\rho^{Bob}_{initial} $ and $\rho^{Bob}_{success}$
have two-dimensional supports, the support of
$\rho^{Bob}_{failure}$ must be a subspace of the support of
$\rho^{Bob}_{success}$. Therefore, we conclude that
$\rho^{Bob}_{success}$ and $\rho^{Bob}_{failure}$ do {\it not}
have orthogonal supports and hence the fidelity
$F({ \rho^{Bob}_{success} \over {\rm Tr}
\rho^{Bob}_{success} }, {\rho^{Bob}_{failure} \over {\rm Tr}
\rho^{Bob}_{failure}} ) \not= 0$.~QED

\section{Some details of
Proof of Lemma~2}
\label{a:prooflemma2}

As before
Alice attaches an ancilla to her
system $A$ and the evolution needed now is
\begin{equation}
|0\rangle_a| j \rangle_A\rightarrow \bigl({1\over{\sqrt{k-1}}}\sum_{i=1;
i\neq  j}^k |i\rangle_a\bigr)| j\rangle_A .
\end{equation}
That is, the state $| j \rangle_A$ of the particle  remains unchanged, 
but the ancilla is brought to an equal superposition of all states
$|1\rangle_a,
\cdots, |k\rangle_a$, with the exception of $|j\rangle_a$.
The evolution of the state of the ancilla and the pair can, therefore,
be summarized as
\begin{eqnarray}
 |0\rangle_a |\Phi_k\rangle &= & |0\rangle_a\bigl({ 1 \over \sqrt{k} }
 \sum_{j=1}^k | j \rangle_A| j \rangle_B\bigr) \nonumber \\
            & \to& {1\over{\sqrt k}}\sum_{i=1}^k |i\rangle_a
 \bigl({1\over{\sqrt{k-1}}} \sum_{j=1;  j\neq i}^k |j\rangle_A|j\rangle_B
\bigr).
\end{eqnarray}
i.e., each state $|i\rangle_a$ of the ancilla is
correlated with a different k-1 
dimensional maximally entangled state.

Next, Alice measures the state of her ancilla.  No matter what result she
obtains, the pair of particles is left in a (k-1)-dimensional maximally 
entangled
state. Which particular state is obtained will depend on Alice's result.
Suppose Alice
finds the ancilla in the state $|i_0\rangle_a$. Then the pair is in the state
${1\over{\sqrt{k-1}}}\sum_{j=1; j\neq i_0}^k|j\rangle_A |j\rangle_B$.  If they
wish, Alice and Bob can now convert this state
into the standard (k-1)-dimensional maximally 
entangled state ${1\over{\sqrt{k-1}}}\sum_{j=1}^{k-1}|j\rangle_A |j\rangle_B$.
This can be realized by Alice communicating to Bob the result of her
measurement, such that both of them know which (k-1)-dimensional maximally
entangled state has
been obtained and then having both of them perform appropriate local
unitary transformations
of their particles.

Now starting with a maximally entangled $r$-dimensional state, one
can repeat our argument to reduce it to a maximally entangled
$(r-1)$-dimensional state, $(r-2)$-dimensional state, etc
until we obtain an $s$-dimensional state. This shows that
any maximally entangled state can be reduced to one with a lower
dimension.

\section{Proof of Lemma~3}
\label{a:prooflemma3}
Since $|\Psi_{\rm pre}^{m,p,0} \rangle$ is
an $m$-ME-state,
all we need to show is the reduction with certainty from
$|\Psi_{\rm pre}^{m,p,q} \rangle $ to $|\Psi_{\rm pre}^{m,p,q-1} \rangle $
whenever $q \geq 1$. The proof here is analogous to that of
Lemma~2.

Suppose Alice attaches an ancilla to her system
and evolves them in the following manner:
\begin{eqnarray}
 |0\rangle_a| j \rangle_A &\rightarrow& \bigl({1\over{\sqrt 
 {p+q}}}\sum_{i=1}^{p+q} |i\rangle_a\bigr)| j \rangle_A, \nonumber \\
& & ~~~{\rm for}~1\leq j\leq
m-p \nonumber \\
|0\rangle_a| j \rangle_A &\rightarrow& \bigl({1\over{\sqrt{p+q-1}}}
\sum_{i=1; i\neq j - (m -p)}^{p+q}
|i\rangle_a\bigr)| j \rangle_A, \nonumber \\
& &
~~~{\rm for}~m-p+1\leq j\leq m+q.
\end{eqnarray}
In words, the ancilla is brought to an equal superposition of all
states $| 1 \rangle_a, \cdots, | p+q \rangle_a$ if the
state of Alice's system is $| j \rangle_A$ where $1 \leq j \leq
m-p$. However, when Alice's system is in $| j \rangle_A$ where
$m-p+1\leq j\leq m+q$, the ancilla is brought to an equal superposition of
all states $| 1 \rangle_a, \cdots, | p+q \rangle_a$ with the exception of
$| j - (m-p) \rangle_a$.
Upon measuring the state of the ancilla and applying
local unitary transformations to their respective systems,
Alice and Bob end up in a new  
precursor $|\Psi_{\rm pre}^{m,p,q-1} \rangle $. This proves the reduction
from $|\Psi_{\rm pre}^{m,p,q} \rangle $ to
$|\Psi_{\rm pre}^{m,p,q-1} \rangle $. By repeating this reduction process,
one can, with certainty, reach $|\Psi_{\rm pre}^{m,p,0} \rangle $
which is an $m$-ME-state.

\section{Proof of Lemma~4}
\label{a:prooflemma4}
It is convenient here to
define $S_{m-r+1} = \sum_{i =m-r +1}^N \lambda_i $.
Then,
\begin{eqnarray}
B^m_{r+1} &>& B^m_r \nonumber \\
{ m \over r+1} [S_{m-r+1}  + \lambda_{m-r} ] &>& { m \over r} S_{m-r+1}
 \nonumber \\
r S_{m-r+1} + r \lambda_{m-r} &>& (r+1) S_{m-r+1}  \nonumber \\
    r \lambda_{m-r}   &>& S_{m-r+1} .
\label{sprime}
\end{eqnarray}
Now,
\begin{eqnarray}
B^m_{r+2} &=& {m \over (r+2)} [ S_{m-r+1} + \lambda_{m-r} + \lambda_{m-r-1}]
          \nonumber \\
          &\geq& {m \over (r+2)} [S_{m-r+1} + 2  \lambda_{m-r}]
          \nonumber \\
          & =& {m \over (r+2)(r+1)} [(r+1)S_{m-r+1} +2 (r+1)\lambda_{m-r}]
              \nonumber \\
         & =& {m \over (r+2)(r+1)} [(r+1)S_{m-r+1} + r \lambda_{m-r} +
          (r+2)\lambda_{m-r} ]  \nonumber \\
          &>& {m \over (r+2)(r+1)} [(r+1)S_{m-r+1} + S_{m-r+1} +
          (r+2)\lambda_{m-r} ]  \nonumber \\
        &=& {m \over (r+2)(r+1)} [(r+2)S_{m-r+1} +
          (r+2)\lambda_{m-r} ]  \nonumber \\
        &=& {m \over (r+1)} [ S_{m-r+1}+
          \lambda_{m-r} ]  \nonumber \\
         &=& B^m_{r+1} ,
\end{eqnarray}
where Eq.\ (\ref{sprime}) is used in obtaining the fifth line. QED.

\section{Proof of Lemma~5}
\label{a:prooflemma5}
Let us consider the list of values of $B^m_1, B^m_2, \cdots , B^m_m$.
Since $ B^m_m=1 > {\rm min}_r  B^m_r$, as a function of $r$, $B^m_r$ must
start to increase at some point. i.e., there exists $r_0$ such that
$B^m_{r_0 +1} > B^m_{r_0 }$. But then, by Lemma~4,
$B^m_{r_0 +2} > B^m_{r_0 +1}$, $B^m_{r_0 +3} > B^m_{r_0 +2}$, etc.
In words, once $B^m_{r}$ starts to increase, it will continue to do so.
Let us focus on the {\it last} minimal point of the function $B^m_r$.
i.e., the {\it largest} value $r_1$ such that $B^m_{r_1} =
{\rm min}_r  B^m_r $. By definition, $B^m_{r_1+1 } > B^m_{r_1}$ which,
from Lemma~4,
implies that
$B^m_{r_1} <
B^m_{r_1 +1 } < \cdots < B^m_m =1$. This completes the first part of
the proof.

Moreover, we claim that $B^m_1 \geq B^m_2 \geq  \cdots \geq B^m_{r_1}$.
We prove this by contradiction. Assuming the contrary, there exists
an $a \leq r_1$ such that $B^m_{a-1} < B^m_{a}$. Then Lemma~4
implies
that $B^m_{ r_1 -1} < B^m_{r_1}$, which is impossible because
it contradicts the fact that $B^m_{r_1}= {\rm min}_r  B^m_r $.

Combining the results of the above two paragraphs, we conclude that
$B^m_1 \geq B^m_2 \geq  \cdots \geq B^m_{r_1} <
B^m_{r_1 +1 } < \cdots < B^m_m =1$. QED.

\section{Proof of Lemma~6}
\label{a:prooflemma6}

The following proves the claim in Lemma~6
that $\lambda'_i = \lambda_i - p/3$.
For simplicity, we shall use projection operators rather than
(POVMs). As noted in Section~\ref{reduction},
there is no loss in generality.
Let $P_1, P_2, \cdots, P_r$ be the set of projection operators
for extracting some 3-ME-state from $\Psi$.

Now suppose $P$ gives a 3-ME-state with a probability $\alpha$.
\begin{equation}
| \Psi \rangle = P | \Psi \rangle + ( 1-P) | \Psi \rangle
\end{equation}
with
\begin{eqnarray}
P | \Psi \rangle &=& \left( \sqrt{\lambda_1} 
P | 1 \rangle \right)  | 1 \rangle  \nonumber \\
& & + \left( \sqrt{\lambda_2} 
P | 2 \rangle \right)  | 2 \rangle  \nonumber \\ 
& & +
\left( \sqrt{\lambda_3} 
P | 3 \rangle \right)  | 3 \rangle .
\end{eqnarray}
Since $P| \Psi \rangle$ is $3$-ME-state with a norm $\alpha$,
its reduced density matrix for $B$,
\begin{equation}
\rho_B = \sum_{i =1}^3 {\alpha  \over 3} | i \rangle \langle i | .
\end{equation}
Equating this with the partial trace of $P | \Psi \rangle \langle  \Psi | P  $
over $H_A$, we find that the
${ \sqrt{\lambda_i} \over { \sqrt{ \alpha \over 3}}}
P | i \rangle $'s form an orthonormal set.
The residual state
\begin{eqnarray}
( 1 -P)| \Psi \rangle &=& \sqrt{\lambda_1 - { \alpha \over  3}} 
 | 1'' 1 \rangle \nonumber \\ & & +
  \sqrt{\lambda_2 - { \alpha \over  3}   } 
     | 2'' 2 \rangle  \nonumber \\ & &
+
   \sqrt{\lambda_3 - { \alpha \over  3}  } 
  | 3'' 3 \rangle .
\end{eqnarray} 
Notice that the $| i'' \rangle$'s are orthonormal because
\begin{eqnarray}
 & & \langle j | (1 - P) (1 -P) | i \rangle  \nonumber \\
&=&  \langle j | (1 - 2P +PP) | i \rangle  \nonumber \\
&=&  \langle j | (1 - 2PP +PP) | i \rangle  \nonumber \\
&=&  \langle j | (1 - PP) | i \rangle  \nonumber \\
&=& 0.
\end{eqnarray}
Here the last equality follows from the fact
that  $
P | i \rangle $'s are orthogonal to one another.
This shows that an extraction of a $3$-ME-state of probability $\alpha$
leads to a decrease of each $\lambda$'s by $\alpha/3 $.
The same argument can be applied to each of $P= P_1, P_2, \cdots, P_r$.
This shows that
after the extraction with a probability
$p$ of a 3-ME-state from $\Psi$, the eigenvalues of the
reduced density matrix of the un-normalized residual state satisfy
$\lambda'_i = \lambda_i - p/3$. QED.


\begin{references}
\bibitem{addremark1} A more precise discussion goes as follows.
Suppose the goal is to transform an initial state $\Psi$ into
$\Upsilon_1$. By keeping track of the results of all
her measurement, Alice will generally obtain the
state of a larger system
$\Upsilon = \Upsilon_1 \otimes  \Upsilon_2$.
However, she can easily reduce its Schmidt coefficients to
those of $ \Upsilon_1$ simply by performing a complete
measurement along the second system.
\bibitem{extension} This symmetry is originally defined only on the subspace
of the Hilbert space with $\lambda_k \not= 0$, but it
can be trivially extended
to the whole Hilbert space by pairing, in the subspace where
$\lambda_k = 0$, {\it any} orthonormal basis $| a''_k \rangle $'s of $H_A$
with {\it any} orthonormal basis $| b''_k \rangle $'s of $H_B$.
\bibitem{max} We use the superscript MAX because, as will be shown in
Section~6, the supremum probability is attainable by the optimal strategy.
\end{references}
\end{document}